\documentstyle[mprocl]{article}

\def\be{\begin{equation}}
\def\ee{\end{equation}}
\def\bea{\begin{eqnarray}}
\def\eea{\end{eqnarray}}
\begin{document}

\title{INFINITE NUCLEAR MATTER ON THE LIGHT FRONT: A MODERN APPROACH TO
  BRUECKNER THEORY}

\author{G. A. MILLER}

\address{Physics Department, Box 351560\\
  University of Washington \\
  Seattle, Washington 98195-1560\\
  E-mail: miller@phys.washington.edu}

\maketitle\abstracts{ 
 Understanding
an important class of experiments   requires  that
light-front dynamics and the related  light
cone variables $k^+,k_\perp$ be used.
 If one uses
 $k^+=k^0+k^3$ as a momentum variable the corresponding canonical
 spatial variable is $x^-=x^0-x^3$ and the time variable is $x^0
 +x^3$. 
This  is  the light front (LF)
approach of Dirac.
A relativistic light front formulation of nuclear
dynamics is developed and applied to treating infinite nuclear matter
in a method which includes the correlations of pairs of nucleons. 
This is light front Brueckner theory.}

\section{Outline}

This talk is divided into four parts.
(1) What is the Light Front Approach? The basic idea is to use a
``time'' variable $c\tau=ct+z$
(2) Why use it? Certain kinds  of high energy experiments are best analyzed
using light front or light cone variables.
(3) Mean field theory results.
(4) Nucleon-nucleon correlations. The way to include these, in any formalism,
is Brueckner Theory.

\section{What is Light Front Dynamics?}
This is a  relativistic treatment of many-body dynamics~\cite{Di 49}
in which the
 ``time'' variable is taken to be \be\tau =ct+z =x^0+x^3\equiv
 x^+.\label{tau}\ee 
 The canonically conjugate 
 ``energy'' variable is $ p^0-p^3\equiv p^-$. 
One of the 
``space'' variables  must be the orthogonal combination
$x^-\equiv t-z$, with its canonically conjugate momentum:
$p^+=p^0+p^3\equiv p^+$. The other variables are
$\vec{x}_\perp,$ and $ \vec{p}_\perp$.

 Our notation is  $A^\pm\equiv A^0\pm A^3$. The point of this was noticed long
 ago by experimentalists. Consider a
particle with a large velocity such that: $\vec v\approx c\hat{e_3}$. In that
case
the momentum  $p^+$
is BIG. An important consequence of using light front variables is that the
usual relation betwen energy and momentum, 
$p^\mu p_\mu =m^2$ becomes
\begin{equation}p^-={1\over p^+}
%\left
(p_\perp^2+m^2),\label{disp}\end{equation}
so that one obtains a relativistic kinetic
energy without a square root 
 operator. Eq.~(\ref{disp}) is of great use in separating
 relative and center-of-mass variables. Another feature is that here the 
vacuum is empty. It contains no virtual-pair states.

\section{Motivation}
It is certainly possible to do quantum mechanics this way, but
why? I think that the use of light front dynamics is mandated if one wants to
correctly understand a large class of high energy nuclear reactions. The most
prominent example is deep inelastic lepton scattering from nuclei.
\subsection{
\bf$x_{Bj}$, Light Front Nuclear Physics and the EMC effect}
Deep inelastic scattering occurs when a quark of four-momentum $q$ strikes a
quark of momentum $p$ that originated from a nucleon of momentum $k$. In that
case, \be x_{Bj}\equiv {-q^2\over 2Mq^0}=
 {p^+\over k^+},\ee for large enough values of $-q^2$ and $q^0$. One studies,
 experimentally and theoretically, the ratio of a cross section(per nucleon)
 $\sigma(A)$  on a nucleus to
 that $\sigma(N)$ on a nucleon. At
 high energies and momentum transfer one might think that
 the 
ratio $\sigma(A)/\sigma(N)$ would be very close to unity.
The European Muon Collaboration  found that this was not so--there is a
depletion (EMC effect)
$\sigma(A)/\sigma(N)\approx 0.85$ in the region $x_{Bj}\approx0.5$
for which valence quark are dominant~\cite{emc}. If there is such a depletion,
and momentum
is conserved, there must be an enhancement of the momenta carried by other
degrees of freedom. This could be manifest  as an enhancement of nuclear pions
for $x_{Bj}\approx 0.1$.
But there were many non-conventional theories of this effect including
swollen nucleus, six-quark cluster, and color conductivity through the entire
nucleus. 
Almost immediately after the EMC effect was
discovered we argued~\cite{bbm}
that another kind of experiment: Drell-Yan production of
muon pairs,  could be used to test the various theories of the EMC effect.
No excess pions were discovered~\cite{dyexp}
and this was termed a crisis in nuclear physics
by Bertsch et al~\cite{missing}.

My opinion is that 
the conventional explanation of nuclear binding and related Fermi
motion effects has never been properly evaluated because of the failure to
re-derive nuclear wave functions using the formalism  (as  given in
reviews~\cite{lcrevs}) of light front
dynamics.
Thus, it has been our  intention to provide realistic and relativistic
calculations of nuclear wave functions using light front
dynamics~\cite{gam97ab}$^-$\cite{mm}.

\subsection{Formal aspects}
To make light front-nuclear physics calculations
we need to know the probability
that a nucleon has a given value of  $k^+$: $f_N(k^+)$. Similarly the
distribution function
for a pion is given by  $f_\pi(k^+)$. I have emphasized deep inelastic
scattering so far,
but these quantities enter into the analyis of 
 many experiments including the (e,e'p) and  (p,pp) reactions~\cite{fs1}.
 The consequence
 of taking $\tau$ of Eq.~(\ref{tau}) as the time variable is that the
 distribution functions $f_{N,\pi}$ are simply related to the absolute
 square of the 
ground state wave function. If ones uses the conventional equal time
formulation, one finds that the same information is encoded in the response
function which involves   matrix elements between the ground
and  an infinite number of excited states.

In light front dynamics,
one only needs the ground state, but one has to obtain this from a consistent
calculation. To illustrate the difficulty one may ask, ``What is $k^+$?''.
Many authors, including myself, have 
 used the idea that $k^+$ is an energy plus momentum to invoke a relation:
 $k^+= M-\epsilon_\alpha =k^3$, where $\epsilon_\alpha$ is an orbital
 binding energy. This relation is not
correct. The variable $k^+$ is a continuous kinematic variable (akin to $k^3$
of the usual quantum mechanics). It is not related to any discrete eigenvalue.

\section{
  Light Front Quantization}

Our motto is that 
we need a ${\cal L}$, no matter how bad! This is necessary in order to derive
expressions for the operators  $P^\pm$
which are the ``momentum'' and ``Hamiltonian'' of the theory.
Consider for example, the Walecka model~\cite{bsjdw} (also called QHD1)
${\cal L}(\phi,V^\mu,\psi)$. The degrees of freedom are 
 nucleon $\psi$, neutral vector meson $V^\mu$, and scalar meson
$\phi$. This is the simplest Lagrangian that can provide even a caricature of
nuclear physics. Exchange of scalar mesons leads to a long ranged attractive
potential and exchange of vector mesons leads to a shorter range and stronger
repulsive
potential. In this way, the nucleons are held together, but are not allowed to
collapse. Given $\cal L$, one constructs the energy-momentum tensor,
$T^{\mu\nu}$. In particular, 
\be P^\mu ={1\over 2}\int d^2x_\perp dx^- T^{+\mu}. \ee
A technical challenge is to 
express $ T^{+\mu}$
in terms of independent variables. For example, the 
nucleon is usually treated as a 4-component spinor. But this particle has spin
1/2, so there are really only two  independent degrees of freedom, denoted as
$\psi_+$. One must express the remaining degrees of freedom in terms of 
$\psi_+$~\cite{lcrevs,gam97ab}. %,gam97b}.

\section{
Infinite Nuclear Matter in  Mean Field Approximation-MFA}

This simple limiting case is the first problem we consider.
The idea behind the mean field approximation is that the sources of mesons
are strong, so there are  many mesons, which can be  treated as classical
fields.
The 
volume if taken as infinite, so that  all positions, and spatial-directions
are equivalent. We treat nuclear matter in its rest frame here.
In that case the solution of the mesonic field equations lead  to the results 
\be
V^\pm=V^0={g_v\over m_v^2}\langle\psi^\dagger(0)\psi(0)\rangle;\quad
 {V}_{i}=0,\label{vf}\ee
\be
\phi={-g_s \over m_s^2}\langle\bar \psi(0)\psi(0)\rangle,\ee
in which the brackets represent ground state matrix elements. The fields
$\phi,V^\pm$  are constants, so the 
nucleon modes are plane waves. One has a  Fermi gas in which
$\psi \sim e^{ik\cdot x}$
and \be i\partial^- \psi_+=g_v\bar{V}^-\psi_++{k_\perp^2+(M+g_s\phi)^2\over
k^+}\psi_+.\label{lfn}\ee The equations (\ref{vf})--(\ref{lfn}) are
 a self-consistent set of equations. 

\subsection{ Nuclear Momentum Content}
One uses the energy-momentum tensor to determine $P^\pm$. One finds
\bea
{P^-\over\Omega}=m_s^2\phi^2+{4\over
(2\pi)^3}\int_F d^2k_\perp dk^+\;{k_\perp^2+(M+g_s\phi)^2\over k^+},\\
{P^+\over\Omega}=m_v^2(V^-)^2+{4\over
(2\pi)^3}\int_F d^2k_\perp dk^+\;k^+,\label{pplus}\eea
in which $\Omega$ is the volume of the system. The Fermi sphere is  determined 
by using an implicit defnition of $k^3$:
\be k^+\equiv \sqrt{(M+g_s\phi)^2+\vec{k}^2} +k^3.\ee Then one may show that
 the energy of the nucleus,   
$E\equiv{1\over 2}\left(P^-+P^+\right)$ is the same~\cite{gam97ab}
as for the  Walecka
model. This is a nice check on the calculation because that model as been
worked
out in a manifestly covariant manner.  Then the minimization,
$\left({\partial (E/A)\over\partial k_F}\right)_\Omega=0$ determines the
value of the Fermi momentum, $k_F$. This very same equation also  
 sets $
P^+=P^-=M_A,$  a most welcome result.

\subsection{Mean field results and implications}
The numerical calculation shows that  
the  LF reproduces standard good
results for energy and  density. But the explicit
decompostion (\ref{pplus}) allows us to determine that 
nucleons carry only 65\% of the nuclear + Momentum ($M_A$) .
A value of  90\% is needed
to explain the EMC effect (in infinite nuclear matter)~\cite{Sick},
so this is a problem.
Furthermore, vector mesons carry a huge  35\% of the + momentum. Because 
 $V^-$ is constant in space-time, $ V^- \ne0$ only if 
 $k^+\to0.$ This means that this effect requires a beam of infinite energy
 to be detected. 
These 
results, which conflict with experiments, might be  artifacts of using
infinite  nuclear matter, or caused by the use of the MFA. More serously, the
${\cal L}$ could be at fault.

\subsection{ Saving Mean   Mean Field Theory?} 
A simple way to improve the phenomenology is to modify $\cal L$, by for example
including 
 scalar meson self coupling terms: $
\phi^3,\phi^4$. A wide variety of parameter sets reproduce the binding
energy and density of nuclear nuclear matter~\cite{fpw87}. For one set,  
nucleons  carry 90\% of $P^+,\;$
so that  vector mesons carry 10\%. This could be acceptable.  of $ P^+$
There is a problem with this parameter set, the related nuclear   
spin-orbit splitting  is found to be too small~\cite{fu96a}.
This is not so bad,
since there are a variety of non-mean field mechanisms which can supply
a spin orbit force. Thus one finds a need to 
 go beyond mean field theory. This involves the introduction of  
light front Brueckner theory. 

\section{Light Front NN interaction}
 The nucleon-nucleon potential is
obtained from one boson exchange using another ${\cal L}$ which includes the
effects of pions and other mesons   absent from QHD1 and in which chiral
symmetry is respected~\cite{mm}.
 The $\tau-$ ordered perturbation theory
rules give expressions~\cite{gam97ab} which can be translated into the usual
language. Schematically, in momentum space we have
\be
V(meson)\sim{1\over q_0^2-\vec{q}^2-\mu^2}\ee
in which $q^\mu$ is the four-mometnum transferred between nucleons and $\mu$ is
the meson mass. This is the standard Yukawa form, except that the effects of
retardation are included via the $q_0$ term. The kernal $\cal K$
is the sum of the 
meson exchanges:
\be{\cal K }=\sum_{meson} V(meson), \ee
in which the mesons are the usual
set of $\pi,\rho,\omega,\sigma, \eta,\delta$.
The potentials are strong so that there effects are taken into account to all
orders by solving the light front version of the Lippman-Schwinger
equation~\cite{fs1}. Schematically we write:
\be
{\cal M}={\cal K}+\int {d^2p_\perp\;d\alpha\over \alpha(1-\alpha)}{\cal K}
{2M^2\over P^2-{p_\perp^2+M^2\over\alpha(1-\alpha)}+i\epsilon}{\cal M},\ee 
in which $P$ is the total four-momentum of the two nucleon system and
$p_\perp,\alpha$ are relative momenta~\cite{gam97ab}.
This equation does not seem
to have
rotational invariance, but this can be recovered by making a change
of variables inwhich the z-component of the relative momentum
is defined implicitly:
\be \alpha\equiv {E(p)+p^3\over2E(p)},\ee
with $E(p)=\sqrt{p_\perp^2+p_3^2+M^2}$.
Then the integrand in the equation above is simplified:
\be
{d^2p_\perp\;d\alpha\over \alpha(1-\alpha)}
{2M^2\over P^2-{p_\perp^2+M^2\over\alpha(1-\alpha)}+i\epsilon}\to
{M^2\over  E(p)} {d^3p\over P^2/4-E^2(p)+i\epsilon}.\ee
This is of the form of the Blankenbecler Sugar equation except that the effects
of retardation must be included.

Given this formalism we followed the usual prescription of varying the
meosn-nucleon  form factors to achieve a reasonably good description of the
data~\cite{mm}. 
\section{
 Light Front Theory of $\infty$ Matter - with  NN Correlations}
I outline our  detailed theory~\cite{mm}. The starting point is a
Lagrangian decomposed into nucleon kinetic terms ${\cal L}_0(N)$, meson kinetic
terms  ${\cal L}_0({\rm mesons})$ and meson-nucleon interactions
${\cal L}_I(N,{\rm mesons})$. Then
\be
{\cal L}={\cal L}_0(N)+{\cal L}_I(N,{\rm mesons})+{\cal L}_0({\rm mesons}) .\ee
The two-nucleon one-boson-exchange-potential OBEP, ${\cal V}(NN)$, does not
enter
so we add it and subtract it:
\be {\cal L}={\cal L}_0(N)-{\cal V}(NN) +
({\cal L}_I(N,{\rm mesons})+{\cal L}_0({\rm mesons})+
{\cal  V}(NN))\label{decomp} \ee %$$
The term in parentheses   
accounts for mesonic content of Fock space. One does perturbation theory in
this operator to learn if one has chosen a nucleon-nucleon potential that is
consistent
with the chosen Lagrangian. 
 The first term of Eq.~(\ref{decomp}) represents the standard
nuclear many body problem. One handles this by introducing the mean field
$MF$:
\be {\cal L}_0(N)-{\cal V}(NN) ={\cal L}_0(N)-U_{MF} +
\left(U_{MF}-{\cal V}(NN)\right)\ee
We choose $U_{MF}$ in the usual way, according to the
independent pair approximation.
In that case the mean field is the folding of scattering  matrix with
                                the nuclear density:
\bea %U_{MF}\sim \left(\rm{Scattering Matrix}_{Pauli}\right) \times \rho\\
 U_{MF}\sim \left(\rm{Brueckner\; G-matrix}\right) \times \rho.\eea

 The result of all of these manipulations is that one obtains a
 full wave function  which contains both nucleon-nucleon
 correlations and explicit
mesons. This procedure is very similar to the usual many-body theory evaluated
with equal time quantization. I stress the differences.
The simplicity of the vacuum allows a relativistic theory to be derived
using non-relativistic techniques. 
We are able to obtain
light front plus-momentum distributions for nucleons  and mesons. The only
technical difference is that we
  include retardation effects 
in our OBEP.
\subsection{Saturation Properties}
We find good results. The binding energy per
nucleon is 14.7 MeV and $k_F=1.37$ Fm.
The compressibility is 180 MeV. Given this, the interesting thing to do is to
assess the influence of this calculation on nuclear structure functions.

\section{ Deep Inelastic Scattering and Drell-Yan Production}
We find $M+g_s\phi=0.79 M$ this is very much larger than the mean field value
of $ 0.56M$. As a result nucleons carry more than  84\% of the nuclear
plus-momentum. The 84\%
 is obtained using only the uncorrelated- Fermi gas part of the wave
function. We also estimate that including the  2p-2h correlations would lead
to nucleons carrrying more than  $90\%$
of the plus momentum. Including 
nucleons with momentum greater than $k_F$ would substantially increase the
computed ratio $F_{2A}/F_{2N}$ because $F_{2N}(x)$ decreases very rapidly
with increasing values of $x$ and because $M^*$ would increase at high momenta.
This is a good start to solving the problems mentioned in the earlier parts of
this talk. Furthermore, we computed the total number of excess pions, and find
that ${N_\pi\over A}=5\%$. This is much smaller than the only
previously computed result~\cite{bf} of  15\%. The quantity $N_\pi$ is not a
direct input into computations, but previous phenomenological
calculations~\cite{jm} 
allow us to hope that  the 
5\% would be consistent with Drell
Yan data. Our present conclusion is that
light front dynamics leads to reasonable nuclear dynamics. The 
90\%, and 5\%
 numbers are an excellent start.

 Clearly, many things remain to be done
with this approach. In the meantime, I would like to emphasize that 
Light Front Nuclear Physics exists! One can use it to understand 
 any high energy nuclear reaction.

\vspace*{-2pt}

\section*{Acknowledgments}
This talk is based on work done with 
 Rupert
Machleidt.
This work was supported in part by the U.S.D.O.E.

\eject


\vspace*{-9pt}

\section*{References}
\begin{thebibliography}{99}
\bibitem{Di 49}  
P.\, A.\, M.\, Dirac, % (1949)} Dirac\, P.\, A.\, M.\, (1949): 
%Forms of relativistic dynamics. 
 {\em Rev. Mod. Phys.} {\bf 21}, 392 (1949).
%%CITATION = RMPHA,21,392;%%
\bibitem{emc}  J. Aubert {\it et al.,}{\em Phys. Lett.} {\bf B123}, 275 (1982).
 R.G. Arnold,  {\it et al.,}{\em Phys. Rev. Lett.} {\bf 52,} 727 (1984).  
A. Bodek,  {\it et al.,}{\em Phys. Rev. Lett.} {\bf 51, } 534 (1983). 
% \bibitem{EMCrevs} S
See  the 
recent  reviews: 
M. Arneodo, {\em Phys. Rep.}
240, 301 (1994). 
D.F. Geesaman, 
K. Saito, A.W. Thomas, Ann. Rev. Nucl. Part. Sci. {\bf 45}, 337 (1995).

\bibitem{bbm} R.P. Bickerstaff, M.C. Birse, G.A. Miller,
{\em Phys. Rev. Lett.} {\bf 53}, 2532 (1984); 
{\em Phys.Rev.} {\bf D33} 3228 (1986). 
\bibitem{dyexp}D.M.  Alde et al. {\em Phys. Rev. Lett.} {\bf 64}, 2479 (1990).
\bibitem{missing} G.F. Bertsch, L. Frankfurt, and M. Strikman,
{\em Science } {\bf 259}, 773 (1993).
\bibitem{lcrevs}
 S. J. Brodsky, H-C Pauli,
S.\,S.\, Pinsky, Phys. Rept. {\bf 301}, 299 (1998); 
%hep-ph/9705477.
%%CITATION = PRPLC,301,299;%%SLAC-PUB-7474, hep-ph/9705477; 
M.  Burkardt Adv. Nucl. Phys. {\bf23}, 1 (1993);
A. Harindranath,
in {\em Light-Front Quantization and Non-Perturbative QCD},
ed. J.P. Vary and F. Wolz (International Institute of Theoretical and Applied
Physics, Ames, 1997).



\bibitem{gam97ab}G.A. Miller, {\em Phys. Rev.} {\bf C56}, R8 (1997);
{\em Phys. Rev.}   {\bf C56} {2789} (1997).
\bibitem{bbm97}
P.G.~Blunden, M.~Burkardt and G.A.~Miller, {\em Phys. Rev.}
{\bf C59}, R2998 (1999); in press {\em Phys. Rev.}
{\bf C},  (1999), nucl-th/9901063.
\bibitem{bm} M.~Burkardt and G.A.~Miller, {\em Phys. Rev.}
  {\bf C58}, 2450 (1998)
\bibitem{mm} G.A. Miller and R. Machleidt {\em Phys. Rev.} {\bf C 60},
035202 (1999);
{\em Phys. Lett.} {\bf B 455},  19 (1999)
.



\bibitem {fs1} L.~L. Frankfurt and M.~I. Strikman 
Phys. Rep. {\bf 76}, 215 (1981).
\bibitem{bsjdw}B.D. Serot and J.D. Walecka, Adv. Nucl. Phys. {\bf 16},1
B.D. Serot and J.D. Walecka,
{\em Int. J. Mod. Phys.} {\bf E6}, 515 (1997).
%%CITATION = IMPAE,E6,515;%%


\bibitem{Sick} I. Sick and D. Day, {\em Phys.Lett.} {\bf B274}, 16 (1992).
    \bibitem{fpw87} R.J.~Furnstahl, C.E.~Price,and G.E.~Walker
      {\em Phys Rev } {\bf C}36, 2590   (1987)
\bibitem{fu96a} 
R.J.~Furnstahl, J.J.~Rusnak and B.D.~Serot,
%``The nuclear spin-orbit force in chiral effective field theories,''
{\em Nucl. Phys.} {\bf A632}, 607 (1998).
%%CITATION = NUPHA,A632,607;%%





\bibitem{bf} B. Friman, V.R. Panharipande and R.B. Wiringa {\em
    Phys. Rev. Lett}
  {\bf 51}, 763 (1983)






 \bibitem{jm} H. Jung and G.A. Miller, Phys. Rev. {\bf C41},659 (1990).

 
\end{thebibliography}
\end{document}